\documentclass
[twocolumn,showpacs,preprintnumbers,prl,final,letterpaper]{revtex4}%
\usepackage{amssymb}
\usepackage{graphicx}
\usepackage{dcolumn}
\usepackage{bm}%
\usepackage{amsmath}%
\usepackage{amsfonts}
\begin{document}
\title{Anisotropic phonon DOS: the application of Rietveld\\and M\"{o}ssbauer texture analysis in aligned powders}
\author{A. I. Rykov,$^{1,2,3}$M. Seto,$^{4}$ Y. Ueda,$^{5}$ and K.~Nomura$^{2}$}
\affiliation{$^{1}$Siberian Synchrotron Radiation Center, Lavrentieva 11, Novosibirsk,
630090, Russia, $^{2}$The University of Tokyo, Hongo 7-3-1, 113-8656, Japan,
$^{3}$Technology Crystals Laboratory "Tecrys", Institutskaya 4/1, 630090,
Novosibirsk, Russia, $^{4}$Research Reactor Institute, Kyoto University, Noda,
Kumatori-machi, Osaka 590-0494, Japan, $^{5}$Institute for Solid State
Physics, University of Tokyo, 5-1-5, Kashiwanoha, Chiba 277-8581, Japan }
\date{\today }

\begin{abstract}
While synthesizing the single crystals of novel materials is not always
feasible, orienting the layered polycrystals becomes an attractive method in
the studies of angular dependencies of inelastic scattering of x-rays or
neutrons. Putting in use the Rietveld analysis of layered structures in novel
manganites and cuprates we develop the studies of their anisotropic properties
with oriented powders instead of single crystals. Densities of phonon states
(DOS) and atomic thermal displacememts (ATD) are anisotropic in the A-site
ordered manganites LnBaMn$_{2}$O$_{y}$ of both $y=5$ and $y=6$ series (Ln=Y,
La, Sm, Gd). We establish the angular dependence of DOS on textures of
arbitrary strengths, link the textures observed by x-ray and $\gamma-$ray
techniques, and solve the problem of disentanglement of Goldanskii-Karyagin
effect (GKE) and texture in M\"{o}ssbauer spectra.

\end{abstract}
\maketitle

\address{$^{1}$ The University of Tokyo, Hongo 7-3-1, 113-8656, Japan\\
$^{2}$ Technology Crystals Laboratory, Institutskaya 4/1, 630090,
Novosibirsk, Russia\\
$^{3}$ Institute for Solid State Physics, University of Tokyo, 5-1-5
Kashiwanoha, Chiba 277-8581, Japan\\
$^{4}$ Research Reactor Institute, Kyoto University, Noda, Kumatori-machi,
Osaka 590-0494, Japan }

Two states of matter traditionally used in structure analysis are single
crystals and polycrystalline powder. When synthesizing the single crystal is a
stiff task the random polycrystalline materials or even polycrystals oriented
on a surface conformably to their persistent crystal habitus can be employed.
Via the method developed initially by Rietveld\cite{Riet} the refinement of
texture, or preferred orientation, is a conventional procedure alongside with
the refinement of atomic structure parameters. Randomization of the powders
favors the perfect refinements of atomic parameters, therefore, any residual
texture is usually not in line with the best sample preparation for x-ray
profile analysis. There appears, however, a class of problems, in which the
well-oriented powders of platy or acicular crystallites can replace the
unavailable single crystals. To them belong the studies in anisotropic
properties of materials, in electric, magnetic properties, and in lattice
dynamics. The texture descriptions and determination of the preferred
orientation parameters are thus becoming the issues of self-sustained interest.

Recently, orienting powders to make the samples with varied degree of
crystallite alignment was suggested\cite{EPL} to be useful in vibrational
spectroscopy of anisotropic materials. When the polycrystalline material is a
ferromagnet or a superconductor it can be thoroughly subjected to texturing in
an external magnetic field.

The vibrational properties in novel manganites and cuprates are anisotropic
owing to their layered structure. Substitution of Fe into $3d$ Jahn-Teller
cation site perturbs the electronic system at small doping rate, and modifies
the existing charge and/or orbital order. However, due to strong
electron-phonon interactions the system keeps its attractiveness for
high-resolution spectroscopies based on nuclear $\gamma-$resonance. The
Fe-doped samples allowed us to probe the anisotropy of the dopant vibrations
due to GKE in $^{57}$Fe M\"{o}ssbauer spectroscopy\cite{RCR}. The same doping
procedure is extremely valuable in the synchrotron radiation vibrational
spectroscopy using the nuclear inelastic scattering (NIS) \cite{Review}. In
this work, we establish the relationships between the parameters of preferred
orientation, the anisotropic phonon DOS, and the M\"{o}ssbauer line
intensities. Using these relationships we propose the novel technique of NIS
on the oriented powder samples.%

\begin{figure}
[ptb]
\begin{center}
\includegraphics[
height=1.6553in,
width=3.2967in
]%
{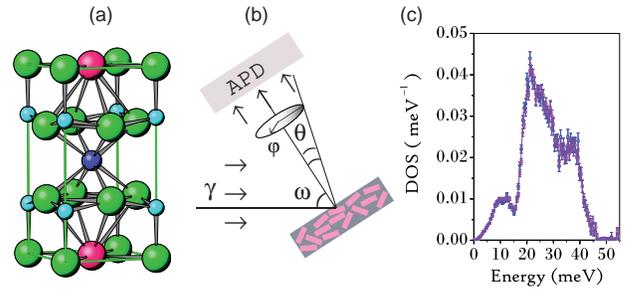}%
\caption{Structure of oxygen-deficient layered oxides, e.g., manganites
LnBaMn$_{2}$O$_{5}$ (a); geometry of nuclear inelastic scattering of
synchrotron radiation (b); phonon density of states of Fe in unoriented powder
of TbBaFe$_{2}$O$_{5}$ (c).}%
\label{f1}%
\end{center}
\end{figure}

Previous works dealt with the texture manifestations in M\"{o}ssbauer
spectra\cite{Pfan1,Pfan2,Gren}. For a unpolarized M\"{o}ssbauer source\ there
occur two parameters of texture, which can be determined from spectra. These
parameters define the so-called "minimum texture function"(MTF). Our approach
is to employ the March-Dollase\cite{Dollase} function (MDF) implemented in
programs for diffraction profile analysis\cite{DBW,Full}. We establish the
relation between Rietveld and M\"{o}ssbauer textures in terms of MDF and MTF,
using the example of quadrupole doublet spectra for the $^{57}$Fe nuclei
occupying $2\%$ of Mn sites in paramagnetic state of novel manganites
LnBaMn$_{2}$O$_{5}$ \cite{Millange} and LnBaMn$_{2}$O$_{6}$%
\cite{NKU,NKYOU,WAR}. The symmetries are tetragonal, except a monoclinic
member YBaMn$_{2}$O$_{6}$, and measurements were made with azimuthal rotation.
Our texture function $T(\theta)$ was uniaxial, $\varphi-$invariant (Fig.1).

In the proposed method, the DOS derivation is based on two NIS patterns and a
Rietveld pattern. Owing to relationship between MTF and MDF, the M\"{o}ssbauer
spectra and Rietveld analysis become mutually complementary techniques.
Indeed, the transmission (M\"{o}ssbauer) and reflection (Bragg or NIS) data
may diverge if there occur some in-depth variations of texture. The $\gamma
-$resonance wavelength (0.86 \AA ) is close to x-ray Mo K$_{a}$(0.7 \AA ), but
the main component of NIS radiation collected by the avalanche photodiode
detector (APD) is Fe $K_{\alpha}$ (1.94 \AA )\cite{Review}, that is close to
x-ray wavelength of Cu $K_{\alpha}$(1.54 \AA ).

In the NIS spectrum of an anisotropic crystal, the phonon DOS is weighted by
squared projection of the phonon polarization vectors to the wave vector of
the x-ray quantum\cite{KCR}. Let the incident beam be launched under the angle
$\vartheta$ with respect to preferred axis ($z$-axis) of a plate- or a
needle-like crystallite. The projected DOS for this crystallite is:%

\begin{equation}
g_{E}(\vartheta)=g_{z}(E)\cos^{2}\vartheta+g_{x}(E)\sin^{2}\vartheta
\end{equation}
To introduce the averaging of the DOS over the ensemble of aligned
crystallites we must integrate these two terms\ with the volume of
crystallites $D(\vartheta,\phi)d\Omega$ whose z-axis lies within the cone
shell element $d\Omega$:%

\begin{equation}
\left\langle g(E)\right\rangle =g_{x}(E)+\Delta g_{zx}(E)\int D(\vartheta
,\phi)\cos^{2}\vartheta d\Omega\label{03}%
\end{equation}
Here $\Delta g_{zx}(E)=g_{z}(E)-g_{x}(E)$ and the orientation distribution
function (ODF) is normalized to unity. The polar ODF $T(\theta)$ is to replace
$D(\vartheta,\phi)$ via the coordinate transform from the frame of the beam to
the frame of the rotation stage. The ratio of angular elements $d\Omega
_{\text{beam}}/d\Omega_{\text{stage}}$ is $\sin$ $\vartheta d\vartheta
d\phi/\sin\theta d\theta d\varphi$ and the Jacobian of this transform is
$\sin$ $\vartheta/\sin\theta$. Using $\cos\vartheta=\cos\theta\cos\omega
-\sin\theta\sin\omega\cos\varphi$ we obtain for the uniaxial symmetry%

\begin{equation}
\left\langle \cos^{2}\vartheta\right\rangle =\left\langle \cos^{2}%
\theta\right\rangle \cos^{2}\omega+%
\frac12
\left\langle \sin^{2}\theta\right\rangle \sin^{2}\omega\label{05}%
\end{equation}
Powder averaging $\left\langle \sin^{2}(\theta)\right\rangle =$ $\smallint
\sin^{2}(\theta)T(\theta)\sin\theta d\theta$ can be expressed via Legendre
function of the first kind $P_{2}(x)=%
\frac12
(3x^{2}-1)$:%

\begin{equation}
\left\langle \sin^{2}\theta\right\rangle =\frac{1}{3}\int_{-\pi/2}^{\pi
/2}T(\theta)\left[  P_{2}(\cos\theta)-1\right]  d(\cos\theta)\label{08}%
\end{equation}
\qquad\qquad

Prior measuring the NIS spectra M\"{o}ssbauer spectroscopy and Rietveld
analysis provide the results in terms of MTF and MDF, respectively. Due to the
axial symmetry, the texture function $T(\theta)$ would have only even terms in
the Legendre expansion series.%

\begin{equation}
T(\theta)=\frac{2n+1}{2}\sum_{n=0}^{\infty}C_{n}P_{n}(\cos\theta)\label{17}%
\end{equation}
The observable quantities, $\left\langle g(E)\right\rangle $, and asymmetries
in M\"{o}ssbauer spectra ($R$ or $\Delta R_{%
\frac12
}$, as denoted below) are fully determined by the integrals of $T(\theta)$,
i.e., (\ref{08}) and $\left\langle \cos^{2}\theta\right\rangle =1-\left\langle
\sin^{2}\theta\right\rangle $, therefore, the function $T(\theta)$ consistent
with experiment is not unique. The simplest function $T(\theta)$ conforming
the Eq. (\ref{08}), is the so-called MTF \cite{Pfan2}, composed of a linear
combination of $P_{0}=1$ and $P_{2}(\cos\theta)$ terms, i.e., the parabolic
function of $\cos\theta$.

Rietveld analysis\cite{Riet,DBW,Full} specifies the preferred orientation with
two fit parameters $r$ and $\Phi$ of the MDF \cite{Dollase}:%
\begin{equation}
M(\theta,\Phi,r)=1-\Phi\left[  1-(r^{2}\cos^{2}\theta+r^{-1}\sin
\theta)^{-\frac{3}{2}}\right] \label{01}%
\end{equation}
%

\begin{figure}
[ptb]
\begin{center}
\includegraphics[
height=2.0392in,
width=3.4575in
]%
{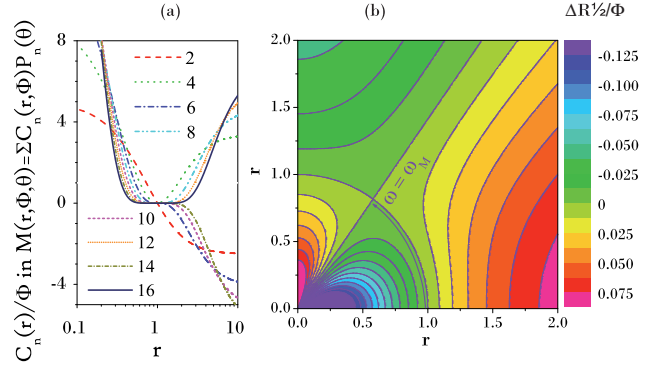}%
\caption{The $r-$dependence of the $\Phi-$scaled coefficients of the Legendre
series expansion of the MDF, Eq. \ref{01} (a) and contour polar plot of the
isotropic ($\alpha=0$) site $\Phi-$scaled spectra asymmetry $\Delta
R_{\frac12 }/\Phi$ in function of the strength of preferred orientation $r$
and incidence angle $\omega$ (b). }%
\label{f2}%
\end{center}
\end{figure}

The angle $\theta$ describes the misfit of the crystallite orientation with
respect to the axis of the uniaxial texture (Fig.1, b), and $\Phi$ is the
fraction of the oriented phase. The March variable $r$ expresses the strength
of the preferred orientation. In a sample made by pressing a layer of plates
(needles) of the initial thickness $d_{0}$ $(d)$ down to the thickness $d$
$(d_{0})$, the degree of compression is $r=d/d_{0}$\cite{Eric}. The function
$M(\theta,\Phi,r)$ conserves the scattering matter, therefore, it has a
clearer physical sense than $T(\theta)$ originally used by Rietveld
$T(\theta)=exp(-r\cdot\theta^{2})$\cite{Riet}. MDF is thus a true normalized
angular distribution, equally good to the textures made of crystallites with
either platy or acicular habitus\cite{Riet,DBW,Full}.

Ericsson and W\"{a}ppling have previously studied in chlorite\cite{Eric} the
effect of texturing induced by compression of various degrees $r_{i}%
=d_{i}/d_{0}$ on the M\"{o}ssbauer line area asymmetry. They have pointed out
that their model "represent an extreme case" and that less randomness were
observed in chlorite samples than predicted by their one-parametric
compression model. Nagy \cite{Nagy} has introduced a second parameter taking
into account the particle shape and assumed that the texturing behavior of the
particles under compression is shape-dependent. For the platy chlorite flakes
Nagy obtained the aspect ratio (cylindrical $h/D$) as large as 0.6
\cite{Nagy}. The March-Dollase function implemented in the FULLPROF
program\cite{Full} is also two-parametric, however, the meaning of the second
parameter $\Phi$ is unrelated to the method of particle alignment. The
unaligned random phase is just added with the fraction of $1-$ $\Phi$. We show
below that the M\"{o}ssbauer line area asymmetry depends for small asymmetries
only on the product $\Phi r$. Difference of physical meanings between $\Phi$
and $r$ manifests itself only in strongly asymmetric spectra.

Measurements with the rotated sample stage inclined to the incident beam
(Fig.1, b) provide us with two components $g_{x}(E)$, $g_{z}(E)$ of uniaxial
DOS in terms of the angle $\omega$ and the parameters $r$ and $\Phi$. Using
MDF (Eq.\ref{01}), we obtain
\begin{align}
\left\langle \sin^{2}\theta\right\rangle  & \equiv V(r,\Phi)=\int_{0}%
^{1}M(\theta,\Phi,r)\sin^{3}\theta d\theta\label{07}\\
V(r,\Phi)  & =\frac{2}{3}(1-\Phi)+\Phi v(r)\\
v(r)  & =\frac{r^{2}}{\varepsilon^{2}(r)}-\frac{\beta(r)}{2\varepsilon^{3}%
(r)}\\
\varepsilon(r)  & =\sqrt{r^{2}-r^{-1}}\\
\beta(r)  & =\ln(2r^{3}+2\sqrt{r^{6}-r^{3}}-1)\text{ \ \ \ }\\
\left\langle \cos^{2}\vartheta\right\rangle  & =\left[  1-V(r,\Phi)\right]
\cos^{2}\omega+\frac{V(r,\Phi)}{2}\sin^{2}\omega
\end{align}
via substitution $\left\langle \sin^{2}\theta\right\rangle $, $\left\langle
\cos^{2}\theta\right\rangle $ into Eq.(\ref{05}). Both $\varepsilon(r)$ and
$\beta(r)$ are imaginary for $0<r<1$, however, $v(r)$ is real for $0<r<\infty
$. The ranges $0<r<1$, and $1<r<\infty$ correspond to platy and stalky habits, respectively.

Now from (\ref{17}) we can express MTF through $r$ and $\Phi$:%
\begin{equation}
\frac{C_{0}}{2}+\frac{5}{2}C_{2}P_{2}(x)=\left(  \frac{15}{4}-\frac{45}%
{4}x^{2}\right)  V(r,\Phi)+\frac{15}{2}x^{2}-\frac{3}{2}\label{09}%
\end{equation}
It is shown in Fig.2 (a) that the Legendre series coefficients of the MDF are
(all, except $C_{0}=1$) returning to zero at $r=1$ and that the MDF is well
approximated by $%
\frac12
C_{0}+(5/2)C_{2}P_{2}(x)$ in some vicinity of random polycrystal $(r=1)$,
because $C_{n}$ $\simeq0$ for $n\geq2$ around $r=1$.%

\begin{figure}
[ptb]
\begin{center}
\includegraphics[
height=2.4163in,
width=3.3503in
]%
{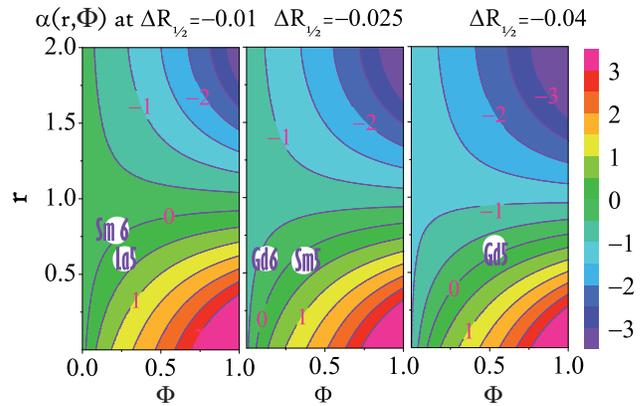}%
\caption{Vibrational anisotropy $\alpha=k^{2}(\langle z^{2}\rangle-\langle
x^{2}\rangle)$ in function of three variables $\alpha(r,\Phi,\Delta
R_{\frac12 })$ , Eq. \ref{15} , is shown as three slices (contour plots) of
$\alpha(r,\Phi)$. In each $\Delta R_{\frac12 }-$matched slice, the oriented
samples of manganites LnBaMn$_{2}$O$_{5}$ and LnBaMn$_{2}$O$_{5}$ are shown by
Ln symbols and oxygen index as spots with coordinates $r,\Phi$ obtained from
Rietveld refinement of parameters of preferred orientation.}%
\label{f3}%
\end{center}
\end{figure}

We are ready for determination of both $g_{x}(E)$ and $g_{z}(E)$. From
Eq.(\ref{03}) a couple of measurements of DOS $g_{1}(E)$ and $g_{2}(E)$ at the
angles $\omega_{1}$ and $\omega_{2}$ leads to:%

\begin{align}
\Delta g_{zx}(E)  & =\frac{\Delta g_{12}(E)}{\left[  1-\frac{3}{2}%
V(r,\Phi)\right]  (\cos^{2}\omega_{1}-\cos^{2}\omega_{2})}\\
g_{x}(E)  & =\overline{g}(E)-\frac{\Delta g_{12}(E)}{2}\frac{(\cos^{2}%
\omega_{1}+\cos^{2}\omega_{2})}{(\cos^{2}\omega_{1}-\cos^{2}\omega_{2})}%
\end{align}
$\overline{g}(E)=g_{1}(E)/2+g_{2}(E)/2$ , $\Delta g_{12}(E)=g_{1}(E)-g_{2}(E)
$.

Now we turn to the asymmetry of M\"{o}ssbauer spectra caused by texture with
parameters $r$ and $\Phi$. First, the intensity ratio for a single crystallite
is to be examined depending on the orientation of the wave vector of the
incident x-ray quantum with respect to the axes of the electric field gradient
(EFG) tensor at the site wherein the $^{57}$Fe nucleus is located. The
$\vartheta$-dependent Clebsch-Gordan coefficients determine the doublet line
intensity ratio $R$\cite{Pfan1,Pfan2}. We found that the quantity, which is
proportional to the amount of the oriented phase $\Phi$ is the deviation of
relative line area from $1/2$:%

\begin{equation}
\Delta R_{%
\frac12
}=\frac{I_{\pm%
\frac12
\rightarrow\pm%
\frac12
}-I_{\pm3/2\rightarrow\pm%
\frac12
}}{2(I_{\pm%
\frac12
\rightarrow\pm%
\frac12
}+I_{\pm3/2\rightarrow\pm%
\frac12
})}=\frac{1}{8}(1-3\cos^{2}\vartheta)\label{11}%
\end{equation}

Powder averaging resolves, using Eq.(\ref{05}), into%

\begin{equation}
\Delta R_{%
\frac12
}=\frac{3}{8}V(r,\Phi)-\frac{1}{4}+\left[  \frac{3}{8}-\frac{9}{16}%
V(r,\Phi)\right]  \sin^{2}\omega\label{16}%
\end{equation}
Eq.(\ref{16}) remains true in $\Phi-$scaled form, i.e. with $\Delta R_{%
\frac12
}/\Phi$ in place of $\Delta R_{%
\frac12
}$, and $v(r)$ in place of $V(r,\Phi)$. From Fig. 2(b) starting with $\Delta
R_{%
\frac12
}/\Phi$ one finds the MDF variable $r$ at any angle $\omega$ except magic
angle $\omega_{M}=54.7^{o}$.%

\begin{figure}
[ptb]
\begin{center}
\includegraphics[
height=2.9879in,
width=3.3589in
]%
{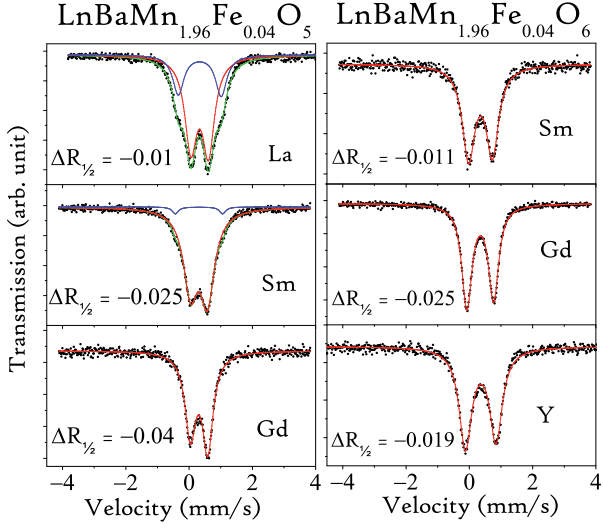}%
\caption{M\"{o}ssbauer spectra in oriented samples of $^{57}$Fe-doped
manganites. Two panels present the oxygen-poor and oxygen-rich series
("O$_{5}$" and "O$_{6}$"). The spectra are fitted either with one or with two
asymmetric doublets. Meltdowns of Mn$^{2+}$/Mn$^{3+}$ and Mn$^{3+}$/Mn$^{4+}%
$charge order in O5 and O6 series, respectively, are seen as single-site
doublets. The occurrence of two sites for Fe in La and Sm members of the
LnBaMn$_{2}$O$_{5}$ series is interpreted as remainder of unmolten charge
order between Mn$^{2+}$ and Mn$^{3+}$. The asymmetries $\Delta R_{\frac12 }$
were fixed to be a single parameter of both doublets. This is because each
crystallite contain both these sites and their $V_{zz}$ direction and sign
coincide according to structure\cite{Millange}.}%
\label{f4}%
\end{center}
\end{figure}

When the measurement is done at $\omega=\omega_{M}$ the texture does not
manifest itself in the spectra. In a simple normal incidence transmission
Mossbauer spectrum $(\omega=0)$, the expression for intensity ratio was
formulated\cite{Pfan1},%

\begin{equation}
R=\frac{\int_{0}^{\pi/2}M(\theta,\Phi,r)(1+\cos^{2}\theta)e^{-\alpha\cos
^{2}\theta}d\theta}{\int_{0}^{\pi/2}M(\theta,\Phi,r)(2/3+\sin^{2}%
\theta)e^{-\alpha\cos^{2}\theta}d\theta}\label{10}%
\end{equation}
however, the solutions were yet found either for texture effects, or for GKE,
separately only. In the proposed synchrotron experiments on anisotropic
powders, both are crucial, therefore, the combined effects of texture and GKE
are of our interest. In (\ref{10}) $\alpha$ is the squared wave vector times
the difference of mean-square vibrational displacements along $V_{zz}$, and in
perpendicular direction, $\alpha=k^{2}(\langle z^{2}\rangle-\langle
x^{2}\rangle)$, $k^{2}=53.35$\AA $^{-2}$. Here again, it is more convenient to
work with $\Delta R_{%
\frac12
}=%
\frac12
(1-R)/(1+R)$. The replacement of $M(\theta,\Phi,r)$ with the MTF, Eq.
\ref{09}, makes our expression for $\Delta R_{%
\frac12
}$ integrable and expressible via two simple sigmoid functions $\sigma
_{0}(\alpha)$ and $\sigma_{1}(\alpha)$:%

\begin{equation}
\Delta R_{%
\frac12
}=\frac{\sigma_{0}(\alpha)+\Phi U(r)\left[  \frac{15}{2}\sigma_{1}%
(\alpha)-5\sigma_{0}(\alpha)-\frac{5}{8}\right]  }{1-20\Phi U(r)\sigma
_{0}(\alpha)}\label{13}%
\end{equation}
with $U(r)=1-3v(r)/2$, $\sigma_{1}(\alpha)=(1+3/2\alpha)\sigma_{0}(\alpha)$, and%

\begin{equation}
\sigma_{0}(\alpha)=\frac{1}{8}-\frac{3}{16\alpha}\left(  1-\frac{e^{-\alpha}%
}{K(\alpha)}\right) \label{12}%
\end{equation}
Here $K(\alpha)=$ $_{1}F_{1}(\frac{1}{2},\frac{3}{2},-\alpha)$ is the Kummer
confluent hypergeometric function. The range of variation for both sigmoid
functions $\sigma_{0}(\alpha)$ and $\sigma_{1}(\alpha)$ is between $-1/4$ and
$+1/8$; for unoriented powder $r=1,$ $\upsilon(1)=0$, and $\Delta R_{%
\frac12
}(\alpha)=$ $\sigma_{0}(\alpha)$ is the exact solution. \ The accuracy of the
solution (\ref{13}) is better than 1\% only in the narrow range of slight
textures ($0.9\lessapprox r\lessapprox1.1$ and $0\leq\Phi\leq1$), in which the
MTF is a perfect approximation for MDF (Eq. \ref{01}). In the same range, the
$\Delta R_{%
\frac12
}(\alpha)$ differs from $\sigma_{0}(\alpha)$ merely by shift of sigmoid
flexpoint:
\begin{equation}
\Delta R_{%
\frac12
}(r,\Phi,\alpha)=\sigma_{0}(\alpha-\alpha_{0}(r,\Phi))\label{14}%
\end{equation}
The approximation (\ref{14}) is as good as (\ref{13}) for $\alpha
_{0}=4.53(1-\Phi r)$. The parameters $r$ and $\Phi$ enter to $\alpha_{0}$ in
equivalent form, however, become nonequivalent if the $\sigma_{0}%
-$approximation (\ref{14}) is extended to a broader range of $r$. We extend it
by taking into account the narrowing sigmoid for stronger textures: $\Delta
R_{%
\frac12
}(r,\Phi,\alpha)=\sigma_{0}(A(r,\Phi)\alpha-B(r,\Phi))$. The accuracy of this
approximation is better than 1\% in the broad range of textures,
$0.3\lessapprox r\lessapprox2$. The $r$ and $\Phi$ dependencies of $A$ and $B$
can be well fitted with $A(r,\Phi)=1+\Phi A_{1}(r)+\Phi^{2}A_{2}(r)+\Phi
^{3}A_{3}(r)$ and $B(r,U)=\Phi B_{1}(r)+\Phi^{2}B_{2}(r)+\Phi^{3}B_{3}(r)$.
The polynomials $A_{i}(r)$ and $B_{i}(r)$ could be expressed as dot products
of coefficients vectors $(a_{0},a_{1},a_{2},a_{3},a_{4})_{i}$ with the vector
$(1,r,r^{2},r^{3},r^{4}) $\ \cite{vectors}. In practice, it is crucial to find
the vibrational anisotropy $\alpha$ starting from $\Delta R_{%
\frac12
}$, therefore, $\alpha$ is given using the function inverse of $\sigma_{0}$:%

\begin{equation}
\alpha(r,\Phi,\Delta R_{%
\frac12
})=\frac{\sigma_{0}^{-1}(\Delta R_{%
\frac12
})}{A(r,\Phi)}+\frac{B(r,\Phi)}{A(r,\Phi)}\label{15}%
\end{equation}

The function of $3$ variables $\alpha(r,\Phi,\Delta R_{%
\frac12
})$ is visualized via several 2D maps (slices, Fig.3). Total range, in which
$\sigma_{0}^{-1}(\Delta R_{%
\frac12
})$ is defined, is $-0.25<\Delta R_{%
\frac12
}<0.125$, and the slices are shown for $\Delta R_{%
\frac12
}=-0.01$, $-0.025$, and $-0.04$. Linear slope $\sigma_{0}^{-1}=30\Delta R_{%
\frac12
}$ is exact near $\Delta R_{%
\frac12
}\simeq0$, but its discrepancy reaches 10\% already at $\Delta R_{%
\frac12
}=-0.05$. The parabolic model $\sigma_{0}^{-1}=31.4\Delta R_{%
\frac12
}+96.5\Delta R_{%
\frac12
}^{2}$ is better than $0.1\%$ in the range $-0.05<\Delta R_{%
\frac12
}<0.05$ that covers the asymmetries observed in our spectra (Fig.4).

We have checked that the rotation of the sample around an axis perpendicular
to the $\overrightarrow{k}-$vector of the incident $\gamma-$quantum induces a
difference of the doublet asymmetry. The same as in Fig.4 absorber of
GdBaMn$_{1.96}$Fe$_{0.04}$O$_{6}$ was installed under the magic angle
$\omega_{\text{M}}=54.7%
{{}^\circ}%
$ and the symmetric doublet was obtained. A very small residual asymmetry can
be explained by the divergence of the incident beam (Fig.5).%

\begin{figure}
[ptb]
\begin{center}
\includegraphics[
natheight=5.110800in,
natwidth=3.715000in,
height=3.5509in,
width=2.5901in
]%
{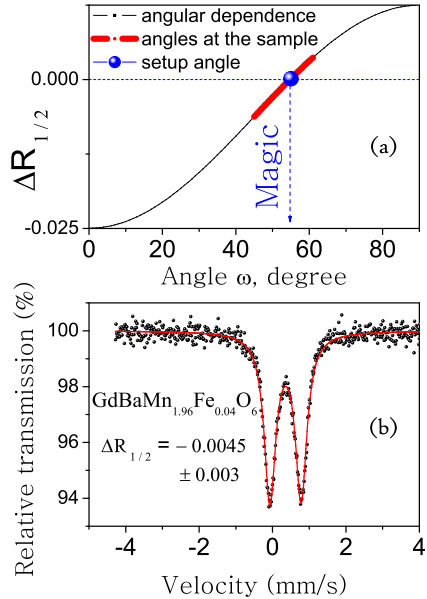}%
\caption{The angular dependence of the doublet asymmetry (a) as given by the
Eq. (17) and the M\"{o}ssbauer spectrum measured under the magic angle
$\omega_{\text{M}}=54.7{{}^\circ}$ between the direction from source to
detector and the normal to the GdBaMn$_{1.96}$Fe$_{0.04}$O$_{6}$ sample
absorber. The distance between source and the sample center was $L=$ 4 cm and
the illuminated area of the sample was 2$\times$2 cm ($d\times d$). The angles
at the absorber surface were thus varied between $\omega_{1}=$ $\omega
_{\text{M}}-\tan^{-1}(\frac{d\cos\omega_{\text{M}}}{L-d\sin\omega_{\text{M}}%
})$ and $\omega_{2}=\omega_{\text{M}}+\tan^{-1}(\frac{d\cos\omega_{\text{M}}%
}{L+d\sin\omega_{\text{M}}})$. This range between $\omega_{1}$ =45${{}^\circ}$
and $\omega_{2}=61{{}^\circ}$ is shown in (a) by bold line. A small residual
asymmetry might have been resulted from the averaging in this range. The
fitted value of $\Delta$R$_{1/2}$ = 0.0045 is 5.5 times smaller than the
normal-incidence value of $\Delta$R$_{1/2}$ in Fig.4. }%
\label{f5}%
\end{center}
\end{figure}

Both oxygen-saturated and oxygen-depleted families of layered manganites show
negative $\Delta R_{%
\frac12
}$, although their left-to-right area ratio changes the sign between "O$_{5}$"
and "O$_{6}$" series. This is because the main axis of EFG $(z)$ is
perpendicular to layers and $V_{zz}>0$ in the pyramid FeO$_{5}$, but
$V_{zz}<0$ in the FeO$_{6}$ octahedron compressed along $z-$axis. Since the
sign of $V_{zz}$ changes from "O$_{5}$" to "O$_{6}$" series the doublet lines
swap their positions. $\Delta R_{%
\frac12
}$ remains negative according to its definition in Eq.(\ref{11}). Therefore,
in both panels of Fig.4, the stronger line is $\pm3/2\rightarrow\pm%
\frac12
$, and the weaker line is $\pm%
\frac12
\rightarrow\pm%
\frac12
$. In this respect, the layered manganites are quite similar to layered
cuprates, wherein the ionic point charge model prescribes the same orientation
and sign of $V_{zz}$\cite{HI}.

The values of $r$ and $\Phi$ for our samples\cite{Y} refined from x-ray
diffraction patterns\cite{Full} are shown by spots in Fig.3, at slices of the
same $\Delta R_{%
\frac12
}$ as shown in Fig.4. The value of $\alpha$ \ increases with increasing size
of Ln within either "O$_{5}$" or "O$_{6}$" series, however, the vibrational
anisotropy is positive in pyramid FeO$_{5}$ ($\langle z^{2}\rangle>\langle
x^{2}\rangle$), but negative in the compressed octahedron FeO$_{6}$ ($\langle
z^{2}\rangle<\langle x^{2}\rangle$). This result is verifiable via refinement
of the factors of anisotropic thermal displacements (ATD) of Mn from neutron
diffraction data, however, till now such the data were refined with $B_{iso}%
-$models\cite{NKYOU,WAR}. On the other hand, in a similar bilayered structure
of YBaFeCuO$_{5}$, the ATD of Fe shaped in FeO$_{5}$ pyramid as prolate
'cigar' was found\cite{Mom}. Also, the oblate 'pancake' is not unexpected
shape of \ ATD in the oblate octahedron of LnBaMn$_{2}$O$_{6}$.

In conclusion, we have shown how the information contained in the x-ray
diffraction full-profile patterns of the aligned powders can be used for
finding the direction-projected components of the phonon DOS. This information
is sufficient, in principle, to solve the problem of determination of two DOS
components from at least two experiments either performed on two samples with
different and non-zero degrees of alignment or conducted on one aligned-powder
sample but under at least two different angles $\omega$. The same information
is contained in assymetry of M\"{o}ssbauer line intensities, however, here the
texture effect is entangled with GKE. Therefore, the useful for DOS or for
other properties information can be disentangled from M\"{o}ssbauer spectra in
only cases when either GKE or ODF are characterized by some separate
experiments. One possibility is to determine the GKE separately from the ATD
refinement using neutron diffraction full-profile analysis superior in the
accuracy of $B_{aniso}$ compared to x-ray Rietveld analysis. Another
possibility is to employ the texture March-Dollase parameters. The latter
possibility was realized in this work for two cases when the sign and
orientation of uniaxial $V_{zz}$ is known, namely, in LnBaMn$_{2}$O$_{5}$
($V_{zz}>0$) and in LnBaMn$_{2}$O$_{6}$ ($V_{zz}<0$). Our method is well
applicable for these tetragonal structures, in which the axis of preferred
orientation coincides with the direction perpendicular to layers and with the
principal axis of electric field gradient. More complicated textured systems,
such as non-uniaxial or magnetic textures were beyond the scope of the present
work. It must be finally emphasized that the scope of possible applications of
the present work is limited by the close correspondence between the
distribution of March-Dollase, which is theoretically substantiated only in
pressed samples\cite{Nagy}, and the real ODF $T(\theta)$, which can be
obtained experimentally with texture goniometers using several analytical
methods, best of which combine the ODF calculation with the Rietveld structure
refinement\cite{combined}. Our approach is also good for samples containing
the randomly oriented phase, because we employed the extended (biparametric)
MDF $M(\theta,\Phi,r)$.

This work was supported by Asahi Glass Foundation and RFBR-JSPS (Grant 07-02-91201).

\section{Figure Captions}

Fig.1. Structure of oxygen-deficient layered oxides, e.g., manganites
LnBaMn$_{2}$O$_{5}$ (a); geometry of nuclear inelastic scattering of
synchrotron radiation (b); phonon density of states of Fe in unoriented powder
of TbBaFe$_{2}$O$_{5}$ (c).

Fig.2. The $r-$dependence of the $\Phi-$scaled coefficients of the Legendre
series expansion of the MDF, Eq. \ref{01} (a) and contour polar plot of the
isotropic ($\alpha=0$) site $\Phi-$scaled spectra asymmetry $\Delta R_{%
\frac12
}/\Phi$ in function of the strength of preferred orientation $r$ and incidence
angle $\omega$ (b).

Fig.3. Vibrational anisotropy $\alpha=k^{2}(\langle z^{2}\rangle-\langle
x^{2}\rangle)$ in function of three variables $\alpha(r,\Phi,\Delta R_{%
\frac12
})$ , Eq. \ref{15} , is shown as three slices (contour plots) of
$\alpha(r,\Phi)$. In each $\Delta R_{%
\frac12
}-$matched slice, the oriented samples of manganites LnBaMn$_{2}$O$_{5}$ and
LnBaMn$_{2}$O$_{5}$ are shown by Ln symbols and oxygen index as spots with
coordinates $r,\Phi$ obtained from Rietveld refinement of parameters of
preferred orientation.

Fig.4. Mossbauer spectra in oriented samples of $^{57}$Fe-doped manganites.
Two panels present the oxygen-poor and oxygen-rich series ("O$_{5}$" and
"O$_{6}$"). The spectra are fitted either with one or with two asymmetric
doublets. Meltdowns of Mn$^{2+}$/Mn$^{3+}$ and Mn$^{3+}$/Mn$^{4+}$charge order
in O5 and O6 series, respectively, are seen as single-site doublets. The
occurrence of two sites for Fe in La and Sm members of the LnBaMn$_{2}$O$_{5}$
series is interpreted as remainder of unmolten charge order between Mn$^{2+}$
and Mn$^{3+}$. The asymmetries $\Delta R_{%
\frac12
}$ were fixed to be a single parameter of both doublets. This is because each
crystallite contain both these sites and their $V_{zz}$ direction and sign
coincide according to structure\cite{Millange}.

Fig.5. The angular dependence of the doublet asymmetry (a) as given by the Eq.
(17) and the M\"{o}ssbauer spectrum measured under the magic angle
$\omega_{\text{M}}=54.7%
{{}^\circ}%
$ between the direction from source to detector and the normal to the
GdBaMn$_{1.96}$Fe$_{0.04}$O$_{6}$ sample absorber. The distance between source
and the sample center was $L=$ 4 cm and the illuminated area of the sample was
2$\times$2 cm ($d\times d$). The angles at the absorber surface were thus
varied between $\omega_{1}=$ $\omega_{\text{M}}-\tan^{-1}(\frac{d\cos
\omega_{\text{M}}}{L-d\sin\omega_{\text{M}}})$ and $\omega_{2}=\omega
_{\text{M}}+\tan^{-1}(\frac{d\cos\omega_{\text{M}}}{L+d\sin\omega_{\text{M}}%
})$. This range between $\omega_{1}$ =45$%
{{}^\circ}%
$ and $\omega_{2}=61%
{{}^\circ}%
$ is shown in (a) by bold line. A small residual asymmetry might have been
resulted from the averaging in this range. The fitted value of $\Delta
$R$_{1/2}$ = 0.0045 is 5.5 times smaller than the normal-incidence value of
$\Delta$R$_{1/2}$ in Fig.4.


\begin{thebibliography}{99}
\bibitem {vectors}Best fits were obtained for the following coefficient
vectors: A$_{1}$=(4.34,-12.3, 12.5,-5.35, 0.84), A$_{2}$=(-5.27, 16.1,-17.9,
8.52,-1.47), A$_{3}$=(2.35,-7.85, 9.47,-4.89, 0.903), B$_{1}$=(11.5,-24.4,
19.7,-8.22, 1.33), B$_{2}$=(-7.79, 24.2,-26.4, 11.8,-1.77), B$_{3}%
$=(8.44,-29.6, 36.5,-18.6, 3.20).
\bibitem {Y}Since the structure of YBaMn$_{2}$O$_{6}$ was monoclinic, the
refinement of texture parameters was not applicable.
\end{thebibliography}
\end{document}